\documentclass[twocolumn,showpacs,preprintnumbers,amsmath,amssymb,floatfix,aps]{revtex4}
\usepackage{graphicx}

\begin{document}


\title{Neutrinos from Gamma-Ray Bursts in Pulsar Wind Bubbles: $\sim 10^{16}\,\,$eV }
\author{Dafne Guetta$^\dagger$ and Jonathan Granot$^\ddagger$}

\affiliation{$^\dagger$Osservatorio astrofisico di Arcetri, 
Fermi 5, 50125 Firenze, Italy; E-mail: 
dafne@arcetri.astro.it\\
$^\ddagger$Institute for Advanced Study, Princeton, NJ
08540; E-mail: granot@ias.edu}

\date{\today}

\begin{abstract}
The supranova model for $\gamma$-ray bursts (GRBs) is becoming increasingly 
more popular.
In this scenario the GRB occurs weeks to years after a 
supernova explosion, and is located inside a pulsar wind bubble (PWB). 
Protons accelerated in the internal shocks that emit the GRB may 
interact with the external PWB photons producing pions which decay into 
$\sim 10^{16}\;$eV neutrinos. A km$^2$ neutrino detector would 
observe several events per year correlated with the GRBs.
\end{abstract}

\maketitle

The leading models for Gamma-Ray Bursts (GRBs) involve a relativistic 
outflow emanating from a compact central source. 
The ultimate energy source is rapid accretion onto a newly formed stellar mass 
black hole. Long duration ($\gtrsim 2\;{\rm s}$) GRBs, which include all GRBs 
with observed afterglow emission and $\sim 2/3$ of all GRBs, are widely 
assumed to originate from a massive star progenitor. 
This is supported by growing evidence for GRBs occurring in star 
forming regions within their host galaxies~\cite{SFR}.
The leading model for long duration GRBs is the collapsar model~\cite{coll}, 
where a massive star promptly collapses into a black hole, 
and forms a relativistic jet that penetrates 
through the stellar envelope and produces the GRB. 
An interesting alternative (though somewhat more debated) model  
is the supranova model~\cite{supra}, where a supernova explosion leaves 
behind a supra-massive neutron star (SMNS), of mass $\sim 2.5-3\,M_\odot$, 
which loses its rotational energy on a time scale, 
$t_{\rm sd}\sim\;$weeks to years, and collapses to a black hole, 
triggering the GRB. The most natural mechanism by which the SMNS can 
lose its rotational energy is through a strong pulsar type wind, 
which is expected to create a pulsar wind bubble (PWB)~\cite{KGGG,GG}.

The prompt $\gamma$-ray emission in GRBs is believed to arise from internal 
shocks within the relativistic outflow, of average isotropic luminosity 
$L\sim 10^{52}\;{\rm erg/s}$, that form due to variability in 
its Lorentz factor, $\Gamma$, on a time scale $t_v$.
These shocks accelerate electrons to relativistic random velocities, which 
then emit synchrotron (and perhaps also synchrotron self-Compton) radiation, 
that constitutes the observed $\gamma$-ray emission. 
In the region where electrons are accelerated, protons are also expected to be 
shock accelerated: the conditions in the dissipation region allow proton 
acceleration up to $\varepsilon_{p,{\rm max}}\sim 10^{20}\;$eV~\cite{Waxman95}.

In this Letter we consider the neutrino production in GRBs that occur
inside PWBs, as is expected in the supranova model for GRBs. 
The neutrino production via photomeson interactions between the relativistic
protons accelerated in the internal shocks and the synchrotron photons that are
emitted in these shocks, has already been calculated~\cite{GRB_nu}, 
while p-p collisions were shown to be relatively unimportant. 
Therefore, we focus on neutrino production through photomeson
interactions with the external photons from the rich radiation field 
inside the PWB. This process turns out to be dominant for a wide range 
of parameters: $t_{\rm sd}\lesssim 0.2\;$yr for a typical
GRB, and $t_{\rm sd}\lesssim 2\;$yr for X-ray flashes.
The neutrinos produced via this mechanism have energies 
$\varepsilon_\nu\sim 10^{15}-10^{17}\,(10^{19})\;$eV for typical GRBs 
(X-ray flashes) and are emitted 
simultaneously with the prompt $\gamma$-ray (X-ray) emission.
Their energy spectrum consists of several power law segments,  
and its overall shape depends on the model parameters.
For 
$t_{\rm sd}\lesssim 0.1\;$yr, 
the $\nu$'s would not be accompanied 
by a detectable GRB.
We find that the $\sim 10^{16}\;$eV neutrino 
fluence from a GRB at $z\sim 0.1-1$ implies $\sim 0.01-1$ upward 
moving muons in a ${\rm km}^{2}$ detector simultaneous with the $\gamma$-rays.
The neutrino signal from an individual GRB that
is pointed towards us, and is therefore detectable in $\gamma$-rays, 
is above the atmospheric neutrino background. 

Recently,~\cite{RMW} calculated the neutrino emission
in the supranova model from protons that escape the internal
shocks region and produce pions inside the SNR shell.

\paragraph*{Neutrino production in the GRB.}---
The internal shocks in GRBs are believed to be mildly 
relativistic. Therefore, the proton energy distribution 
should be close to that for Fermi acceleration
in a Newtonian shock, $dn_p/d\varepsilon_p\propto\varepsilon_p^{-2}$. 
Moreover, the power law index of the electron and proton energy spectra 
should be the same, and the values inferred for the electrons
(from the prompt GRB spectrum) are $dn_e/d\varepsilon_e\propto\varepsilon_e^{-p}$ 
with $p\sim 2-2.5$. We therefore adopt 
$dn_p/d\varepsilon_p\propto\varepsilon_p^{-2}$.
Primed (un-primed) quantities are measured in the comoving (lab) frame.
Protons of energy $\varepsilon_p$ interact mostly with photons 
that satisfy the $\Delta$ resonance condition, 
$\varepsilon_{p,\Delta}=0.3\,{\rm GeV^2}/\varepsilon_\gamma$. 
The minimal photon energy relevant for p$\gamma$ 
interactions is 
$\varepsilon_{\gamma,{\rm min}}=
0.3\,{\rm GeV^2}/\varepsilon_{p,{\rm max}}\sim 
3\times 10^{-3}\;$eV.
For reasonable parameters 
[$t_{\rm sd,-1}=t_{\rm sd}/(0.1\,{\rm yr})\gtrsim 0.26$]
this falls above the self absorption frequency~\cite{GG},
and the relevant part of the PWB spectrum
consists of two power laws:
$dn_\gamma/d\varepsilon_\gamma\propto\varepsilon_\gamma^{-3/2}$ for 
$\varepsilon_\gamma<\varepsilon_{\gamma b}=h\nu_{bm}$ and  
$\propto\varepsilon_\gamma^{-s/2-1}$
for $\varepsilon_\gamma>\varepsilon_{\gamma b}$, where 
$\nu_{bm}\approx 1.6 \times 10^{15} t_{sd,-1}^{-3/2}$ Hz
is the peak frequency of the PWB $\nu F_{\nu}$ spectrum,
$n_\gamma$ is the photon number density, and $s\approx 2.2$ is the power 
law index of the PWB electrons. 
Photons of energy $\varepsilon_{\gamma b}$ satisfy the $\Delta$
resonance condition with protons of energy
\begin{equation} 
\varepsilon_{pb}=4.4\times 10^{16}{\xi_{e/3}^2\beta_{b,-1}^{3/2}
t_{\rm sd,-1}^{3/2}\over\eta_{2/3}^{5/2}\epsilon_{be/3}^{2}
\epsilon_{bB,-3}^{1/2}E_{53}^{1/2}\gamma_{w,4.5}^{2}}\;{\rm eV}\ ,
\end{equation}
where $\beta_{b}c=0.1\beta_{b,-1}c$ is the 
velocity of the SNR shell, 
$\gamma_{w}=10^{4.5}\gamma_{w,4.5}$ and $E_{\rm rot}=10^{53}E_{53}\;$erg
are the Lorentz factor and total energy of the pulsar wind, 
$\xi_{e}=\xi_{e/3}/3$ ($\eta=\eta_{2/3}2/3$)
is the fraction of the wind energy in $e^\pm$ pairs (protons), and
$\epsilon_{be}=\epsilon_{be/3}/3$ ($\epsilon_{bB}=10^{-3}\epsilon_{bB,-3}$)
is the fraction of PWB energy in electrons (magnetic field). 
The corresponding neutrino energy is 
$\varepsilon_{\nu b}\approx\varepsilon_{pb}/20
\sim 2\times 10^{15}t_{\rm sd,-1}^{3/2}\;{\rm eV}$,
similar to that expected from the p$\gamma$ 
interactions with the GRB photons~\cite{GRB_nu}, for 
$t_{\rm sd}\sim 0.1\;$yr. 
The GRB emission can be detected simultaneously with these neutrinos 
only if the Thompson optical depth is $\lesssim 1$, which can be obtained 
for $t_{\rm sd}\sim 0.1\;$yr only if the SNR shell is clumpy~\cite{GG}. 
In the case of a uniform shell we need $t_{\rm sd}\gtrsim 0.4\;$yr,
and in turn $\varepsilon_{\nu b}\gtrsim 2\times 10^{16}\;$eV.

The internal shocks occur over a distance $\Delta R\sim R=2\Gamma^2 c t_v$.
Thus, the optical depth to photo-pion production at the $\Delta$ resonance,
for protons of energy $\varepsilon_p$, is
\begin{equation}\label{tau_pg_GRB} 
\tau_{p\gamma}= 
\sigma_{p\gamma}\varepsilon_\gamma{dn_\gamma\over d\varepsilon_\gamma}R=
{1.0\xi_{e/3}^3E_{53}^{1/2}
\Gamma_{2.5}^2 t_{v,-2}(\varepsilon_p/\varepsilon_{pb})^\beta\over 
f_{1/3}^{2}\eta_{2/3}^{5/2}\epsilon_{be/3}^{5/2}
\gamma_{p,4.5}^{2}\beta_{b,-1}^{1/2}t_{\rm sd,-1}^{3/2}}
\end{equation}
where $\Gamma_{2.5}=\Gamma/10^{2.5}$, $t_v=10^{-2}t_{v,-2}\;$s,
$\sigma_{p\gamma}\approx 0.5\,{\rm mb}$,
$\varepsilon_\gamma=0.3\,{\rm Gev^2}/\varepsilon_p$, 
$f=f_{1/3}/3$ is the fraction of the PWB radius up to which
most of its radiation is emitted,
and $\beta=s/2$ ($1/2$) for $\varepsilon_p<\varepsilon_{pb}$ 
($\varepsilon_p>\varepsilon_{pb}$) is the spectral slope of the 
seed PWB synchrotron 
photons.
The fraction of the proton energy that is lost to pion
production is~\cite{GGnuPWB} 
\begin{equation}\label{f_pg_GRB}
f_{p\pi}(\varepsilon_p)\approx 1-\exp\left[-\tau_{p\gamma}(\varepsilon_p)/5\right]
\approx\min\left[1,\tau_{p\gamma}(\varepsilon_p)/5\right]\ .
\end{equation}
The factor 5 is since the proton
loses $\sim 0.2$ of its energy in a single interaction.
We denote $\varepsilon_{p}$ for which $f_{p\pi}(\varepsilon_{p})\approx 1$ 
by $\varepsilon_{p\tau}=10^{18}\varepsilon_{p\tau18}\;$eV 
[i.e. $\tau_{p\gamma}(\varepsilon_{p\tau})\equiv 5$], and obtain
\begin{equation}\label{e_tau}
\varepsilon_{p\tau18}=\left\{\begin{matrix}
{0.20f_{1/3}^{20/11}\eta_{2/3}^{25/11}
\epsilon_{be/3}^{3/11}\beta_{b,-1}^{43/22}t_{\rm sd,-1}^{63/22}
\over\xi_{e/3}^{8/11}\epsilon_{bB,-3}^{1/2}E_{53}^{21/22}
\gamma_{p,4.5}^{2/11}\Gamma_{2.5}^{20/11}t_{v,-2}^{10/11}}
& {\varepsilon_{p\tau}\over\varepsilon_{pb}}<1\cr & \cr
{1.2f_{1/3}^{4}\eta_{2/3}^{5/2}\epsilon_{be/3}^{3}
\gamma_{p,4.5}^{2}\beta_{b,-1}^{5/2}t_{\rm sd,-1}^{9/2}
\over \xi_{e/3}^{4}\epsilon_{bB,-3}^{1/2}E_{53}^{3/2}
\Gamma_{2.5}^{4}t_{v,-2}^{2}}
& {\varepsilon_{p\tau}\over\varepsilon_{pb}}>1 \end{matrix}\right. \ ,
\end{equation}

The decay of charged pions created in interactions between PWB photons and 
GRB protons, produces high energy neutrinos, $\pi^+\rightarrow \mu^+ +\nu_{\mu}
\rightarrow e^+ +\nu_e +\bar{\nu}_{\mu} +\nu_{\mu}$, where each neutrino 
receives $\sim 5\%$ of the proton energy.

The energy of the protons accelerated in the 
internal shocks of GRBs is expected to
be similar to the $\gamma$-ray energy output in the GRB~\cite{Waxman95}. 
This implies a $\nu_\mu$ fluence,
\begin{eqnarray}\nonumber
f_{\nu_\mu}=f_{0}f_{p\nu}\ \ \ , \ \ \ f_{0}={E_{\gamma,iso}\over 32\pi d_L^2}
=1\times 10^{-5}{E_{\gamma,53}\over d_{L28}^{2}}\;{\rm erg\over cm^{2}}\ ,
\\ \label{F_nu_GRB}
f_{p\nu}={\int d\varepsilon_p(dN_p/d\varepsilon_p)\varepsilon_p 
f_{p\nu}(\varepsilon_p)\over
\int d\varepsilon_p(dN_p/d\varepsilon_p)\varepsilon_p}\ ,
\quad\quad\quad\quad\quad
\end{eqnarray}
where $E_{\gamma,{\rm iso}}=10^{53}E_{\gamma,53}\;$erg is the isotropic equivalent 
energy in $\gamma$-rays, 
$f_{p\nu}(\varepsilon_p)=f_{p\pi}(\varepsilon_p)f_{\pi\nu}(\varepsilon_p)$,
while $f_{p\pi}(\varepsilon_p)$ is given in Eq. (\ref{f_pg_GRB}) and 
$f_{\pi\nu}(\varepsilon_p)$ is the fraction of the original pion energy,
$\varepsilon_\pi\approx 0.2\varepsilon_p$,
that remains 
when it decays. We have taken into account that
although the initial neutrino flavor ratio 
$\Phi_{\nu_e}:\Phi_{\nu_\mu}:\Phi_{\nu_\tau}$ is $1:2:0$, 
neutrino oscillations cause it
to be $1:1:1$ at the Earth.

\begin{figure}[!t]
\includegraphics[width=3.3in]{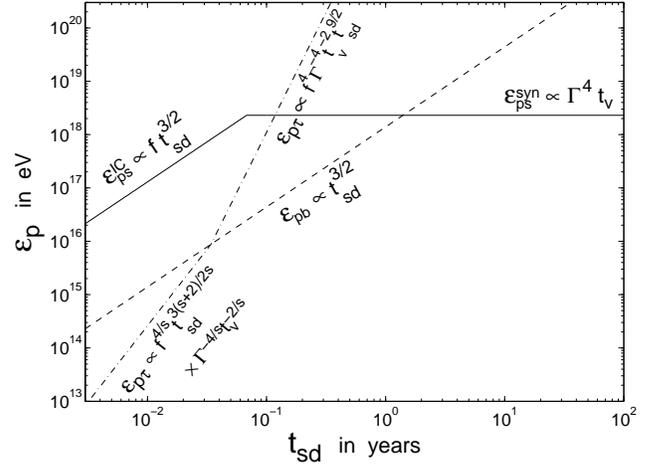}
\caption{\label{fig1}
The proton energies that correspond to break energies in the 
neutrino spectrum, $\varepsilon_\nu\approx\varepsilon_p/20$, 
as a function of $t_{\rm sd}$. 
}
\end{figure}

The pions may lose some energy via synchrotron or inverse-Compton 
(IC) emission. If these energy losses are important, then the final 
energy left in the $\pi^+$ when it decays, out of which $3/4$ goes 
to neutrinos, can be much smaller than its original energy. We find 
that $f_{\pi\nu}(\varepsilon_p)\approx 1-\exp(-t'_{\rm rad}/\tau'_\pi)
\approx \min(1,t'_{\rm rad}/\tau'_\pi)$, where
$\tau'_\pi\approx 2.6\times 10^{-8}\varepsilon'_\pi/(m_\pi c^2)\;$s 
is the life time of the pion, and 
$t'_{\rm rad}=(t_{\rm syn}^{\prime -1}+t_{\rm IC}^{\prime -1})^{-1}
\approx\min(t'_{\rm syn},t'_{\rm IC})$ is the time for radiative losses
due to both synchrotron and IC losses, the time for which are 
$t'_{\rm syn}$ and $t'_{\rm IC}$, respectively. We have
\begin{equation}\label{t_rad}
t'_{\rm syn}={3m_\pi^4c^3\over 4\sigma_T m_e^2 \varepsilon'_\pi U'_B}\quad ,\quad
t'_{\rm IC}={3m_\pi^4c^3\over 
4\sigma_T m_e^2 \varepsilon'_\pi U'_\gamma(\varepsilon'_\pi)}\ ,
\end{equation}
where $U'_B={B'}^2/8\pi$ is the energy density in the magnetic field, within the 
shocked fluid behind the internal shocks, and $U'_\gamma(\varepsilon'_\pi)$ is the 
energy density of photons below the Klein-Nishina limit, 
$\varepsilon'_{\gamma,KN}=(m_\pi c^2)^2/\varepsilon'_\pi$.
IC losses due to scattering of the GRB photons were shown to be 
unimportant~\cite{GRB_nu}. We consider the IC losses due to the 
upscattering of the external PWB photons, and find 
\begin{eqnarray}\label{t_syn}
{t'_{\rm syn}/\tau'_\pi}= 0.21\,\epsilon_e\epsilon_B^{-1}
L_{52}^{-1}\Gamma_{2.5}^8 t_{v,-2}^2\varepsilon_{\pi18}^{-2}
\ ,\quad\quad\quad\quad\quad\quad\ 
\\ \label{t_IC}
{t'_{\rm IC}/\tau'_\pi}= 0.67f_{1/3}^2\xi_{e/3}^{-1}
\epsilon_{be/3}^{1/2}\epsilon_{bB,-3}^{-1/2}E_{53}^{-1}
\beta_{b,-1}^{2}t_{\rm sd,-1}^3\varepsilon_{\pi18}^{-2}\,\, ,\,
\end{eqnarray}
where $\epsilon_e$ and $\epsilon_B$ 
are the equipartition parameters of the GRB, and 
$\varepsilon_{\pi18}=\varepsilon_{\pi}/10^{18}\;$eV. 
The radiative losses become important for $t'_{\rm rad}<\tau'_\pi$, 
which corresponds to $\varepsilon_{p}>\varepsilon_{ps}=
\min(\varepsilon_{ps}^{\rm syn},\varepsilon_{ps}^{\rm IC})
\approx 5\varepsilon_{\pi s}\approx 20\varepsilon_{\nu s}$, where
\begin{eqnarray}\label{e_syn}
\varepsilon_{ps18}^{\rm syn}= 2.3\,\epsilon_e^{1/2}
\epsilon_B^{-1/2}L_{52}^{-1/2}\Gamma_{2.5}^4 t_{v,-2}\ ,
\quad\quad\quad\quad\quad\ \ \, 
\\ \label{e_IC}
\varepsilon_{ps18}^{\rm IC}= 4.1f_{1/3}\xi_{e/3}^{-1/2}
\epsilon_{be/3}^{1/4}\epsilon_{bB,-3}^{-1/4}E_{53}^{-1/2}
\beta_{b,-1}t_{\rm sd,-1}^{3/2}\ .\ 
\end{eqnarray}
The protons may also lose energy via p$\gamma$ interactions with
the GRB photons~\cite{GRB_nu}, However, $\tau_{p\gamma}$ for this process
is typically $\lesssim 1$, so that it does not have a large effect on 
p$\gamma$ interactions with the PWB photons, on which we focus.

\begin{figure}[!t]
\includegraphics[width=3.3in]{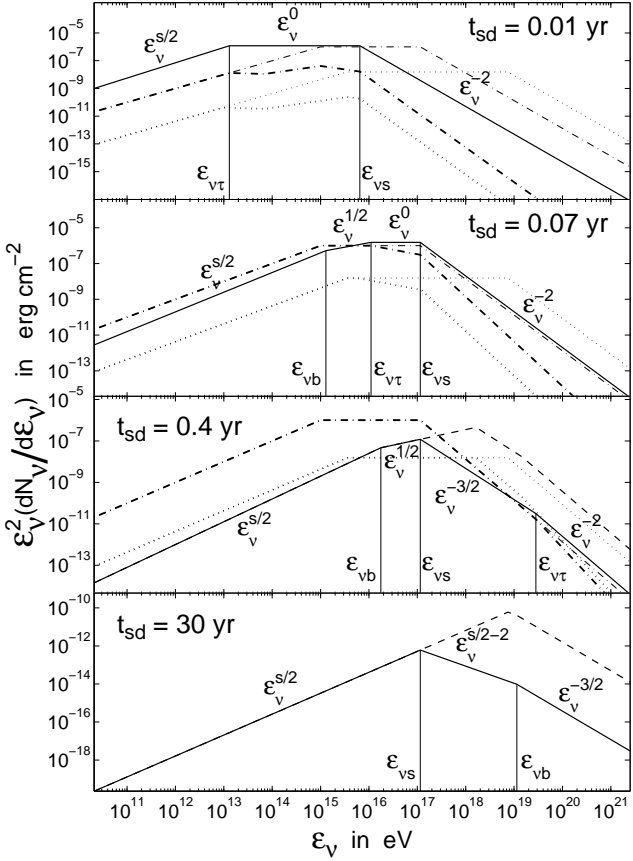}
\caption{\label{fig2}
The muon neutrino spectrum, 
for our fiducial parameters and $t_{\rm sd}=0.01,\, 0.07,\, 0.4,\, 30\;$yr,
for $(\Gamma,t_{v,-2})=(10^{2.5},1)$ ({\it solid}),
which correspond to the 4 different orderings of the break energies,
and $(\Gamma,t_{v,-2})=(600,5)$ ({\it dashed}).
For comparison we show the spectrum of~\cite{GRB_nu}
for  $(\Gamma,t_{v,-2})=(10^{2.5},1)$ ({\it dot-dashed}) and $(600,5)$
({\it dotted}), where the thick (thin)
lines are with (without) the effects of the PWB radiation field 
(that inflict energy losses on the protons, pions and muons).}
\end{figure}

\begin{figure}[!t]
\includegraphics[width=3.3in]{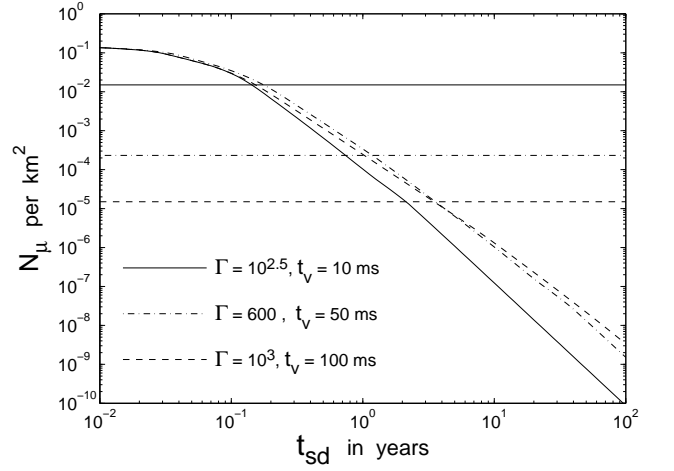}
\caption{\label{fig3}
The number of muon events per GRB in a ${\rm km^2}$ detector,
as a function of $t_{\rm sd}$, for or fiducial parameters and 
three different values of $(\Gamma,t_v)$.
The horizontal lines show the events expected
due to p$\gamma$ interactions with the GRB photons~\cite{GRB_nu}.}
\end{figure}

Since the life time of the muons is $\sim 100$ times longer than that of the pions,
they would have significant radiative losses at an energy of 
$\varepsilon_{\mu s}\sim\varepsilon_{\pi s}/10\approx\varepsilon_{ps}/50$.
This causes a reduction of up to a factor of $3$ in the total neutrino flux
in the range $\sim (0.1-1)\varepsilon_{\nu s}$,
since only $\nu_\mu$ that are produced directly in $\pi^+$ decay contribute 
considerably to the neutrino flux.
Note that since both ratios in Eqs. (\ref{t_syn}) and (\ref{t_IC}) 
scale as $\varepsilon_{\pi}^{-2}$, we always have
$t'_{\rm rad}/\tau'_\pi\propto\varepsilon_{\pi}^{-2}$, 
and therefore the spectrum 
steepens by a factor of $(\varepsilon_\nu/\varepsilon_{\nu s})^{-2}$ for 
$\varepsilon_\nu>\varepsilon_{\nu s}$. This is also evident from the fact that
$f_{\pi\nu}(\varepsilon_p\approx 5\varepsilon_\pi)
\approx\min(1,t'_{\rm syn}/\tau'_\pi,t'_{\rm IC}/\tau'_\pi)$.

Fig. \ref{fig1} shows the proton energies corresponding to the
neutrino break energies $\varepsilon_{\nu s}$, $\varepsilon_{\nu b}$ and 
$\varepsilon_{\nu\tau}\approx\varepsilon_{p\tau}/20$, as a function of $t_{\rm sd}$. 
From Eqs. (\ref{tau_pg_GRB}), (\ref{e_syn}) and (\ref{e_IC}),
we see that $\tau_{p\gamma}\propto\Gamma^2 t_{\rm sd}^{-3/2}$, 
$\varepsilon_{ps}^{\rm syn}\propto\Gamma^4$ and 
$\varepsilon_{ps}^{\rm IC}\propto t_{\rm sd}^{3/2}$.
For a fixed value of $t_{\rm sd,-1}=1$,  $\Gamma_{2.5}\gtrsim 1$ implies 
$\tau_{p\gamma}\gtrsim 1$ and 
$\varepsilon_{ps}=\varepsilon_{ps}^{\rm IC}={\rm const}\sim 10^{18}\;$eV,
while $\Gamma_{2.5}\lesssim 1$ implies 
$\varepsilon_{ps}=\varepsilon_{ps}^{\rm syn}\propto\Gamma^4$.
This implies a stronger neutrino emission which can reach higher energies
for larger values of $\Gamma$. Since a larger $\Gamma$ implies a lower typical 
synchrotron frequency for the prompt GRB, this may apply to X-ray flashes,
if they are indeed GRBs with relatively large Lorentz factors $\Gamma$ and/or 
a large $t_v$~\cite{XRF}.

As can be seen from Fig. \ref{fig1}, for relevant parameters,
there are four different orderings of these break energies:
(i) $\varepsilon_{\nu\tau}<\varepsilon_{\nu b}<\varepsilon_{\nu s}$,
(ii) $\varepsilon_{\nu b}<\varepsilon_{\nu\tau}<\varepsilon_{\nu s}$,
(iii) $\varepsilon_{\nu b}<\varepsilon_{\nu s}<\varepsilon_{\nu\tau}$,
(iv) $\varepsilon_{\nu s}<\varepsilon_{\nu b}<\varepsilon_{\nu\tau}$.
Each of these orderings results in a different shape for the spectrum,
that consists of 3 or 4 power laws, as can be seen in Fig. 2 (solid line).
The peak of the $\varepsilon_\nu^2(dN_\nu/d\varepsilon_\nu)$ spectrum,
in units of $f_0$, is
\begin{equation}\label{g}
\begin{matrix}
{(4+s)/4s+\ln(\varepsilon_{\nu s}/\varepsilon_{\nu\tau})\over
\left[{(4+s)/2s}+
\ln\left({\varepsilon_{\nu s}/\varepsilon_{\nu\tau}}\right)\right]^{2}} 
& \ \ \varepsilon_{\nu\tau}<\varepsilon_{\nu b}<\varepsilon_{\nu s}\cr & \cr
{5/4+\ln(\varepsilon_{\nu s}/\varepsilon_{\nu\tau})\over
\left[5/2+\ln(\varepsilon_{\nu s}/\varepsilon_{\nu\tau})\right]^{2}} 
& \ \ \varepsilon_{\nu b}<\varepsilon_{\nu\tau}<\varepsilon_{\nu s}\cr & \cr
(3/16)\sqrt{\varepsilon_{\nu s}/\varepsilon_{\nu\tau}}
& \ \  \varepsilon_{\nu b}<\varepsilon_{\nu s}<\varepsilon_{\nu\tau}\cr & \cr
{s(4-s)\over 16}
(\varepsilon_{\nu s}/\varepsilon_{\nu b})^{s/2}
\sqrt{\varepsilon_{\nu b}/\varepsilon_{\nu\tau}} 
& \ \  \varepsilon_{\nu s}<\varepsilon_{\nu b}<\varepsilon_{\nu\tau}\end{matrix}\ .
\end{equation}
For example, spectrum (iii) applies for $t_{\rm sd}= 0.4\,$yr,
where the GRB is detectable, and is given by
\begin{equation}\label{spectrum3}
{\varepsilon_\nu^2{dN_\nu\over d\varepsilon_\nu}/f_0
\over
{3\over 16}\sqrt{\varepsilon_{\nu s}\over\varepsilon_{\nu\tau}}}=
\left\{\begin{matrix}
\left({\varepsilon_{\nu b}\over\varepsilon_{\nu s}}\right)^{1\over 2}
\left({\varepsilon_{\nu}\over\varepsilon_{\nu b}}\right)^{s\over 2}
& \ \ \varepsilon_{\nu}<\varepsilon_{\nu b}\cr & \cr
(\varepsilon_{\nu}/\varepsilon_{\nu s})^{1/2} 
& \ \ \varepsilon_{\nu b}<\varepsilon_{\nu}<\varepsilon_{\nu s}\cr & \cr
(\varepsilon_{\nu}/\varepsilon_{\nu s})^{-3/2} 
& \ \  \varepsilon_{\nu s}<\varepsilon_{\nu}<\varepsilon_{\nu\tau}\cr & \cr
\left({\varepsilon_{\nu s}\over\varepsilon_{\nu\tau}}\right)^{3\over 2}
\left({\varepsilon_{\nu\tau}\over\varepsilon_{\nu}}\right)^2
& \ \  \varepsilon_{\nu}>\varepsilon_{\nu\tau}\end{matrix}\right. \, .
\end{equation}
Fig. \ref{fig2} shows the muon neutrino spectrum for our fiducial 
parameters (solid line). The spectrum of the other neutrino flavors
is the same, while the spectrum of the $\gamma$-rays from the pion decay 
is almost the  same, just substituting 
$\varepsilon_\gamma\approx 2\varepsilon_\nu$ and with a 
normalization larger by a factor of 4.
In Fig. \ref{fig2} we also compare our neutrino spectra with
that of~\cite{GRB_nu} for different internal shock parameters.
We see that for reasonably large values of $\Gamma$ and $t_v$,
the PWB photons are the dominant target for photomeson interactions
for $t_{\rm sd}\lesssim 4\;$yr.
The afterglow neutrino emission with 
$\varepsilon_{\nu}\gtrsim 10^{17}\;$eV~\cite{nuafter} should be  
distinguishable from the prompt 
emission (from p$\gamma$ interaction with either GRB or PWB photons) 
due to the very different spectrum, and may also be delayed in time.

\paragraph*{Implications}\hspace{-0.23cm}---
The high energy, $\sim 10^{16}$ eV, neutrinos produced by photomeson 
interactions between protons accelerated in the internal shocks of 
GRBs and the PWB photons may be
detected by the future neutrino telescopes~\cite{Gaisser}.
The probability, $P_{\nu\mu}$, that a neutrino would produce a high
energy muon in the detector is approximately given by the ratio
of the high energy muon range to the neutrino mean free path.
For the neutrinos studied here, $P_{\nu\mu}\approx 1.3\times 10^{-3}
(\varepsilon_\nu/10^{3}\,{\rm TeV})^{\beta}$, with $\beta=1$ for
$\varepsilon_\nu<10^3$ TeV and $\beta=1/2$ for
$\varepsilon_\nu>10^3$ TeV~\cite{Gaisser}. 

In Fig.  \ref{fig3} we report the expected number of events in a
km$^2$ detector as a function of $t_{\rm sd}$, for our fiducial
parameters, for a GRB at $z\sim 1$ with $E_{\gamma,{\rm iso}}=10^{53}$ ergs.
We consider 
three different sets of values for $\Gamma$ and $t_v$,
and compare the resulting numbers with the expected number of events due
to photomeson interactions with the GRB photons~\cite{GRB_nu}.
For a typical GRB with  $\Gamma=10^{2.5}$ and $t_v=10\;$ms, $\nu$'s 
from interactions with PWB photons dominate over $\nu$'s from
GRB photons for $t_{\rm sd}\lesssim 0.2\;$yr.

GRBs that occur inside spherical PWBs with $t_{\rm sd}\sim 0.1-0.2\;$yr would
have a peculiar and short lived afterglow emission~\cite{GG}.
Since they occur for a wide range of $t_{\rm sd}$ values, we expect their
rate to be similar to that of typical GRBs (i.e. $\sim 10^{3}$
yr$^{-1}$ beamed toward us). For these values of $t_{\rm sd}$ we expect 
$\sim 0.01-0.1$ events per burst corresponding to the detection of several tens
of neutrino induced muons per year.
These neutrino bursts should be easily detected above the
background, since the neutrinos would be correlated, both in time and
direction, with the GRB $\gamma$-rays. Note, that at the high energy
considered, knowledge of burst direction and time will allow to
discriminate the neutrino signal from the background by looking not
only for upward moving neutrino induced muons, but also by looking for
down-going muons.
For larger values of $0.2\;{\rm yr}\lesssim t_{\rm sd}\lesssim 1\;$yr,
the neutrino flux due to photomeson interaction with the PWB photons
will dominate over the one due photomeson interaction with the GRB
photons~\cite{GRB_nu}, if the GRB has $\Gamma\gtrsim 600$ and
$t_v\gtrsim 50\;$ms. Since larger $\Gamma$ and $t_v$ imply a lower typical 
synchrotron frequency for the prompt GRB, this may apply to X-ray flashes,
if they are indeed GRBs with relatively large Lorentz factors and/or a large 
variability time, $t_v$~\cite{XRF}. For these events no afterglow emission 
has been detected and this can be explained considering the fact that for
$t_{\rm sd}\lesssim 1$ yr the GRB would have a peculiar and short lived
afterglow emission. The typical neutrino energy is expected to be in the range 
$\sim 10^{15}-10^{17}\;$eV for typical GRBs, and $\sim 10^{15}-10^{19}\;$eV 
for X-ray flashes.

We expect $10^{-5}-0.01$ events per X-ray flash for this
range of $t_{\rm sd}$ corresponding to a detection of $0.01-10$ events per
year. Again, these neutrino bursts should be easily detected above the
background, since the neutrinos would be correlated, both in time and
direction, with the X-rays of the X-ray flashes.
This neutrino emission would be simultaneous with the
$\gamma$-ray/X-ray emission from the GRB and should be easily
distinguishable from neutrinos emitted after the $\gamma$-ray 
phase of the GRB~\cite{nuafter} or $\lesssim 100$ s before the 
GRB~\cite{nuprec}.
Detection of high energy neutrinos will test the shock acceleration
mechanism and the suggestion that GRBs are the sources of ultra-high
energy protons, since $\gtrsim 10^{16}$ eV neutrino production requires
protons of energy $\gtrsim 10^{18}$ eV, and will help  
to establish whether indeed most GRBs occur inside PWBs.

\paragraph*{Acknowledgments.}

We thank J.N. Bahcall, A. Loeb and P. M\'esz\'aros  for useful discussions.
JG is supported by the Institute for Advanced Study, 
funds for natural sciences. 
DG thanks the IAS in Princeton for the hospitality and the 
nice working atmosphere during this work. 

\end{document}